\documentclass[12pt]{article}
\usepackage{amsmath,latexsym}
\usepackage{graphicx}
\usepackage{amsmath}
\usepackage{amsfonts}
\usepackage{amssymb}%
\begin{document}

\title{{Wormholes with low energy density}}
   \author{
Peter K. F. Kuhfittig*\\  \footnote{kuhfitti@msoe.edu}
 \small Department of Mathematics, Milwaukee School of
Engineering,\\
\small Milwaukee, Wisconsin 53202-3109, USA}

\date{}
 \maketitle

\begin{abstract}\noindent
In spite of their speculative nature, traversable 
wormholes are a topic of interest that started 
with the Einstein-Rosen bridge in 1935 and became
a major research area with the introduction of the 
Morris-Thorne wormhole in 1988.  It also became apparent 
in time that such wormholes are likely to be compact 
stellar objects, akin to neutron stars.  Although 
widely discussed, wormholes having a low energy 
density may therefore not be massive enough to 
exist on a macroscopic scale.  Important examples 
are wormholes based on a noncommutative-geometry 
background and wormholes supported by the  negative 
energy density sourced by the Casimir effect.  The 
main goal of this paper is to invoke $f(Q)$ 
modified gravity to provide the extra degrees of 
freedom to help overcome these obstacles.
\\
Keywords
\\ Low-Energy-Density Wormholes,
    Noncommutative Geometry, Casimir Effect, 
    $f(Q)$ gravity

\end{abstract}

\section{Introduction}\label{E:introduction}
Wormholes are handles or tunnels in spacetime
connecting widely separated regions of our
Universe or different universes altogether.  
While wormholes may be as good a prediction 
of Einstein's theory as black holes, they are 
subject to severe restrictions from quantum 
field theory, calling for the existence of 
``exotic matter" \cite{MT88}.  This violation 
is more of a practical than conceptual problem, 
as illustrated by the Casimir effect \cite{hC48}:
exotic matter can be made in the laboratory.
Being a rather small effect, the practical 
challenge lies in generating and sustaining 
a negative energy density of sufficient 
magnitude and volume to support a macroscopic 
object.  In this paper we discuss the more 
general problem of wormholes having a low 
energy density by invoking $f(Q)$ gravity, a
fairly recent modification of Einstein's theory. 

\section{Background}\label{S:background}

\subsection{Morris-Thorne wormholes}
In 1988, Morris and Thorne \cite{MT88} proposed 
the following static and spherically symmetric
line element for a wormhole spacetime, possibly 
motivated by the original Schwarzschild solution:
\begin{equation}\label{E:line1}
ds^{2}=-e^{2\Phi(r)}dt^{2}+\frac{dr^2}{1-b(r)/r}
+r^{2}(d\theta^{2}+\text{sin}^{2}\theta\,
d\phi^{2}),
\end{equation}
using units in which $c=G=1$.  Here $\Phi=\Phi(r)$ 
is called the \emph{redshift function}, which must 
be finite everywhere to prevent the occurrence of 
an event horizon. The function $b=b(r)$ is called 
the \emph{shape function} since it determines the
spatial shape of the wormhole when viewed, for
example, in an embedding diagram.  The spherical 
surface $r=r_0$ is called the \emph{throat} of 
the wormhole.  According to Ref. \cite{MT88}, at 
the throat, $b=b(r)$ must satisfy the following 
conditions: $b(r_0)=r_0$, $b(r)<r$ for $r>r_0$, and 
$b'(r_0)<1$, called the \emph{flare-out condition}.
This condition can only be met by violating the 
null energy condition (NEC), which states that
\begin{equation}
   T_{\alpha\beta}k^{\alpha}k^{\beta}\ge 0
\end{equation}
for all null vectors $k^{\alpha}$, where
$T_{\alpha\beta}$ is the energy momentum tensor.
Matter that violates the NEC is called ``exotic"
in Ref. \cite{MT88}.  In particular, for the 
radial outgoing null vector $(1,1,0,0)$, the 
violation reads $T_{\alpha\beta}k^{\alpha}k^{\beta}
=\rho+p_r<0$.  Here $T^t_{\phantom{tt}t}=-\rho(r)$
is the energy density, $T^r_{\phantom{rr}r}=
p_r(r)$ is the radial pressure, and
$T^\theta_{\phantom{\theta\theta}\theta}=
T^\phi_{\phantom{\phi\phi}\phi}=p_t(r)$
is the lateral (transverse) pressure.  Our
final requirement is \emph{asymptotic flatness:}
$\text{lim}_{r\rightarrow\infty}\Phi(r)=0$ and
$\text{lim}_{r\rightarrow\infty}b(r)/r=0$.

For later reference, we now state  the Einstein
field equations:
\begin{equation}\label{E:E1}
  \rho(r)=\frac{1}{8\pi}\frac{b'}{r^2},
\end{equation}
\begin{equation}\label{E:E2}
   p_r(r)=\frac{1}{8\pi}\left[-\frac{b}{r^3}+
   2\left(1-\frac{b}{r}\right)\frac{\Phi'}{r}
   \right],
\end{equation}
and
\begin{equation}\label{E:E3}
   p_t(r)=\frac{1}{8\pi}\left(1-\frac{b}{r}
   \right)
   \left[\Phi''-\frac{b'r-b}{2r(r-b)}\Phi'
   +(\Phi')^2+\frac{\Phi'}{r}-
   \frac{b'r-b}{2r^2(r-b)}\right].
\end{equation}

\subsection{Wormholes with low energy density}
\label{SS:low}
As noted in the Abstract, wormholes are likely 
to be compact stellar objects, but the original 
formulation of Morris-Thorne wormholes \cite
{MT88} does not call for any such requirement.  
To appreciate the problem, let us recall that 
noncommutative geometry, an offshoot of string 
theory, can in principle support traversable 
wormholes.  This is based on the realization 
that coordinates may become noncommutative 
operators on a $D$-brane \cite{eW96, SW99}.  
A critical feature based on the uncertainty 
principle is that noncommutativity replaces
point-like particles by smeared objects
\cite{SS03, NSS06, NS10}.  The idea is to 
eliminate the divergences that normally occur
in general relativity.  As discussed in Ref.
\cite{NSS06}, this objective can be met
by showing that spacetime can be encoded in the
commutator $[\textbf{x}^{\mu},\textbf{x}^{\nu}]
=i\theta^{\mu\nu}$, where $\theta^{\mu\nu}$ is
an antisymmetric matrix that determines the
fundamental cell discretization of spacetime
in the same sense that Planck's constant $\hbar$
discretizes phase space.  According to Refs.
\cite{NM08, LL12}, a relatively simple way to 
model the smearing is by means of the so-called 
Lorentzian distribution of minimal length 
$\sqrt{\gamma}$ instead of the commonly employed 
Dirac delta function, i.e., we replace the 
point-like Dirac delta function by the following
smooth distribution: the energy density of
a static and spherically symmetric and
particle-like gravitational source is
given by
\begin{equation}\label{E:rho}
  \rho (r)=\frac{m\sqrt{\gamma}}
     {\pi^2(r^2+\gamma)^2}.
\end{equation}
The usual interpretation is that the
gravitational source causes the mass $m$ of
a particle to be diffused throughout the
region of linear dimension $\sqrt{\gamma}$
due to the uncertainty.

Based on these considerations, it is not 
immediately obvious how one can determine the 
size and mass of the wormhole.  So we first 
need to lay the groundwork by following Ref. 
\cite{NSS06}.  Here it is pointed out that 
it is possible to implement the noncommutative 
effects in the Einstein field equations $G_{\mu\nu}
=\frac{8\pi G}{c^4}T_{\mu\nu}$ by modifying
only the energy momentum tensor, while leaving 
the Einstein tensor $G_{\mu\nu}$ intact.  It is 
emphasized in Ref. \cite{NSS06} that the 
noncommutative-geometry background is an 
intrinsic property of spacetime rather than 
some kind of superimposed structure.  So this 
has a direct effect on the mass-energy and 
momentum distributions.  The concomitant
determination of the spacetime curvature then
explains why the Einstein tensor can be left
unchanged.  As a consequence, when describing 
a wormhole, both the length scales and mass 
can be macroscopic, to be confirmed in Sec. 
\ref{S:throat}.  Moreover, noncommutative-geometry 
wormholes based on the Casimir effect are 
discussed in Ref. \cite{pK23a}.  Both are 
examples of wormholes with a low energy density.

\subsection{$f(Q)$ gravity}
Attempts to overcome the theoretical and practical 
problems confronting Morris-Thorne wormholes have 
relied heavily on various modified gravitational 
theories.  A recently proposed modified theory, 
called $f(Q)$ gravity, is due to Jimenez, et al. 
\cite{JHK18}.  Here $Q$ is the non-metricity 
scalar from the field of differential geometry.  
The action for this gravitational theory is
\begin{equation}\label{E:action}
   S=\int \frac{1}{2}f(Q)\sqrt{-g}\,d^4x
   +\int\mathcal{L}_m\sqrt{-g}\,dx^4,
\end{equation}
where $f(Q)$ is an arbitrary function of $Q$, 
$\mathcal{L}_m$ is the Lagrangian density of 
matter, and $g$ is the determinant of the metric 
tensor $g_{\mu\nu}$.  Even though it is a fairly 
new theory, numerous applications have already 
been found; see, for example, Refs.  \cite{JHK18, 
zH22, zH21, gM21, uS21, aB21, gM22, fP22, HMSS, 
lH23}.  This topic will be discussed further in 
Sec. \ref{S:f(Q)}.

\section{Throat size and mass}
   \label{S:throat}
In this section, we will determine both the 
throat size and mass of the wormhole.  So our 
first task is to obtain the shape function from 
Eqs. (\ref{E:E1}) and (\ref{E:rho}), previously 
discussed in Ref. \cite{pK23}.

\begin{multline}\label{E:shape}
   b(r)=\int^r_{r_0}8\pi(r')^2\rho(r')dr'+r_0\\
   =\frac{4m}{\pi}
  \left[\text{tan}^{-1}\frac{r}{\sqrt{\gamma}}
  -\sqrt{\gamma}\frac{r}{r^2+\gamma}-
  \text{tan}^{-1}\frac{r_0}{\sqrt{\gamma}}
  +\sqrt{\gamma}\frac{r_0}{r_0^2
  +\gamma}\right]+r_0\\
   =\frac{4m}{\pi}\frac{1}{r}
  \left[r\,\text{tan}^{-1}\frac{r}{\sqrt{\gamma}}
  -\sqrt{\gamma}\frac{r^2}{r^2+\gamma}-
  r\,\text{tan}^{-1}\frac{r_0}{\sqrt{\gamma}}
  +\sqrt{\gamma}\frac{r_0r}{r_0^2
  +\gamma}\right]+r_0.
\end{multline}
Observe that $\text{lim}_{r\rightarrow
\infty}b(r)/r=0$; to ensure asymptotic
flatness, we retain the assumption
$\text{lim}_{r\rightarrow\infty}\Phi(r)=0$.
Here we can simply let $B=b/\sqrt{\gamma}$ 
be the form of the shape function even 
though $B(r_0)\neq r_0$.  The reason is 
that $B$ can be rewritten as a function of
$r/\sqrt{\gamma}$:
\begin{multline}\label{E:shape}
   \frac{1}{\sqrt{\gamma}}
   \,b(r)=
   B\left(\frac{r}{\sqrt{\gamma}}\right)=\\
   \frac{1}{\sqrt{\gamma}}\frac{4m}{\pi}
   \left (\frac{\sqrt{\gamma}}{r}\right )
   \left[\frac{r}{\sqrt{\gamma}}
   \,\text{tan}^{-1}\frac{r}{\sqrt{\gamma}}
   -\frac{\left(\frac{r}{\sqrt{\gamma}}\right)^2}
   {\left(\frac{r}{\sqrt{\gamma}}\right)^2+1}
  -\frac{r}{\sqrt{\gamma}}\,
  \text{tan}^{-1}\frac{r_0}{\sqrt{\gamma}}
  +\frac{r}{\sqrt{\gamma}}
  \frac{\frac{r_0}{\sqrt{\gamma}}}
  {\left(\frac{r_0}{\sqrt{\gamma}}\right)^2+1}
  \right]+\frac{r_0}{\sqrt{\gamma}}.
\end{multline}
Observe that
\begin{equation}\label{E:throat}
   B\left(\frac{r_0}{\sqrt{\gamma}}\right)
   =\frac{r_0}{\sqrt{\gamma}},
\end{equation}
the analogue of $b(r_0)=r_0$.  Since $B$
is a function of $r$, we may consider the
line element
\begin{equation}\label{E:line2}
ds^{2}=-e^{2\Phi(r)}dt^{2}+\frac{dr^2}
{1-\frac{B(r/\sqrt{\gamma})}{r/\sqrt{\gamma}}}
+r^{2}(d\theta^{2}+\text{sin}^{2}\theta\,
d\phi^{2}).
\end{equation}
It now becomes apparent that in view of
Eq. (\ref{E:throat}), this line element
represents a wormhole with throat radius
$r_0/\sqrt{\gamma}$, while retaining
asymptotic flatness.  It is shown in Ref. 
\cite{pK23} that the flare-out condition 
is met.  Given that $\gamma$ is a small 
constant, it also follows that $r_0/
\sqrt\gamma$ is macroscopic.  Similar 
comments can be made about wormholes 
whose energy violation is due to the Casimir 
effect, whose energy density is given by 
$-\hbar c\pi^2/720r^4$.   

We are now in a position to estimate the mass 
$m(r)$ of the wormhole.  From Eq. (\ref{E:rho}), 
\begin{equation}
   m(r)=\int^r_{r_0}\rho(r')4\pi (r')^2\,dr'
   =\frac{2m}{\pi}
  \left[\text{tan}^{-1}\frac{r}{\sqrt{\gamma}}
  -\frac{r\sqrt{\gamma}}{r^2+\gamma}
   -\text{tan}^{-1}\frac{r_0}{\sqrt{\gamma}}
  +\frac{r_0\sqrt{\gamma}}{r_0^2+\gamma}
    \right].
\end{equation}
Since $m$ in Eq. (\ref{E:rho}) represents the 
mass of a particle, we conclude that the mass 
$m(r)$ cannot be very large.  It has been shown, 
however, that Morris-Thorne wormholes are
actually compact stellar objects \cite{pK22}.  
The implication is that noncommutative-geometry
inspired and Casimir wormholes are likely to be 
microscopic after all. This will be discussed 
further in the next two sections.

\section{Inflating Lorentzian wormholes}
Before continuing, let us briefly consider the
question of inflating Lorentzian wormholes.
It has been suggested that wormholes of the 
Morris-Thorne type may actually exist on 
microscopic scales and that a sufficiently 
far advanced civilization may therefore be 
able to enlarge such a wormhole to macroscopic 
size.  This possibility was explored in Ref. 
\cite{tR92} by assuming that the wormhole is 
embedded in a flat de Sitter space.  Assuming
that $\Phi'(r)\equiv 0$, the time-dependent 
inflationary background is given by
\begin{equation}
ds^{2}=-dt^{2}+e^{2\chi t}\left[\frac{dr^2}{1-b(r)/r}
+r^{2}(d\theta^{2}+\text{sin}^{2}\theta\,
d\phi^{2})\right],
\end{equation} 
where $\chi=\sqrt{\Lambda/3}$ and $\Lambda$ is 
the cosmological constant.  Suppose we now have 
two observers situated on opposite sides of the 
wormhole throat and separated by an initial 
proper distance $l_0$ at $t=0$.  If $l(T)$ is 
the separation at the end of inflation at 
$t=T$, then, according to Ref. \cite{tR92},
\begin{equation}\label{E:Planck}
   l_0<\frac{e^{-\chi T}}{\chi}.
\end{equation} 
This yields $l_0<10^{-67} \text{cm}
\ll 10^{-33} \text{cm},$ the Planck length. 
Since the Planck length is usually regarded 
as the smallest distance that makes physical 
sense, Condition (\ref{E:Planck}) cannot be
met.  Similarly, an initially Planck-sized 
wormhole would be enlarged enormously and 
could even exceed our present cosmological 
horizon.  So inflation alone could not give 
rise to macroscopic wormholes.

\section{Wormholes in $f(Q)$ gravity}
    \label{S:f(Q)}
Following the discussion in Ref. \cite
{aB21}, the non-metricity scalar $Q$ for line 
element (\ref{E:line1}) is given by
\begin{equation}
   Q=-\frac{b}{r^2}\left[
   \frac{rb'-b}{r(r-b)}+\phi'\right]
\end{equation}     
and the field equations are 
\begin{multline}\label{E:Einstein1}
   8\pi\rho(r)=\frac{1}{2r^2}\left(
   1-\frac{b}{r}\right)\left[2rf_{QQ}Q'
   \frac{b}{r-b}+\right. \\ \left.f_Q
   \left(\frac{b}{r-b}(2+r\phi')+
   \frac{(2r-b)(rb'-b)}{(r-b)^2}\right)  
   +f\frac{r^3}{r-b}\right],
\end{multline}         
\begin{multline}\label{E:Einstein2}    
   8\pi p_r(r)=-\frac{1}{2r^2}\left(
   1-\frac{b}{r}\right)\left[2rf_{QQ}Q'
   \frac{b}{r-b}+\right. \\ \left.f_Q
   \left(\frac{b}{r-b}\left(2+\frac{rb'-b}{r-b}
   +r\phi'\right)-2r\phi'\right) 
   +f\frac{r^3}{r-b}\right],  
\end{multline}     
\begin{multline}\label{E:Einstein3}
  8\pi p_t(r)=-\frac{1}{4r}\left(
   1-\frac{b}{r}\right)\left[-2r\phi'f_{QQ}Q'
   +\phantom{\frac{b}{r-b}}\right. \\ \left.f_Q
   \left(2\phi'\frac{2b-r}{r-b}-r(\phi')^2+
   \frac{rb'-b}{r(r-b)}
   \left(\frac{2r}{r-b}
   +r\phi'\right)-2r\phi''\right) 
   +2f\frac{r^2}{r-b}\right], 
\end{multline} 
where $f=f(Q)$, $f_Q=\frac{df(Q)}{dQ}$, and 
$f_{QQ}=\frac{d^2f(Q)}{dQ^2}$.  To keep the 
analysis tractable, we will again follow 
Ref. \cite{aB21} and assume that $\phi'(r)
\equiv 0$, also called the ``zero-tidel-force 
solution."  

Our next step is to check the null energy
condition given the $f(Q)$-gravity background.  
Using Eqs. (\ref{E:Einstein1}) and 
(\ref{E:Einstein2}), we get
\begin{equation}
  \rho(r_0)+p_r(r_0)=\frac{1}{8\pi}f_Q
  \frac{1}{r_0^2}[b'(r_0)-1]<0
\end{equation}
due to the flare-out condition.  So the NEC 
is indeed violated, as long as $f_Q$ is positive. 

Since we are primarily interested in qualitative 
results, our focus is necessarily more narrow.  
Returning to the non-metricity scalar $Q$, it was 
noted after Eq. (\ref{E:action}) that $f(Q)$ is 
an arbitrary function of $Q$.  This gives us 
considerable leeway in choosing $f(Q)$, starting 
with the highly idealized form $f(Q)=\alpha Q
+\beta$ \cite{zH22}, where $\alpha$ and $\beta$ 
constants, also discussed in Ref. \cite{pK23}.  
This form has one serious drawback however: since 
$f''(Q)=0$ and $f'(Q)=$ a constant, this produces 
the Einstein field equations with a cosmological 
constant \cite{lH23}.  Since $f(Q)$ is arbitrary, 
we could simply choose $f(Q)=\alpha Q^{1+\epsilon}
+\beta$, $0<\epsilon \ll 1$.  This form is 
arbitrarily close to $f(Q)=\alpha Q +\beta$, 
while avoiding the above drawbacks.  In other 
words, we can view $f(Q)=\alpha Q +\beta$     
as a convenient approximation that is sufficient 
for our purposes, as we will confirm below. 
(From now on, we assume that $\alpha$ is positive 
to ensure that $f_Q$ is also positive.)

A preferred approach, proposed in this paper, is
to use the form $f(Q)= cQ+de^{\alpha Q}$.  Here 
$\alpha>0$ is assumed to be sufficiently small 
for the form $f(Q)= cQ+de^{\alpha Q}$ to become 
an adequate approximation for the linear form 
$f(Q)=cQ+d$.  This will also allow us to use 
the simpler form 
\begin{equation}
   f(Q)=de^{\alpha Q},
\end{equation}
as can be readily shown: returning to Eq. 
(\ref{E:Einstein1}) with $\Phi'(r)\equiv 0$, 
we obtain 
\begin{multline}
8\pi\rho(r)=\frac{1}{2r^3}(r-b)
   \left[2rf_{QQ}Q'
   \frac{b}{r-b}+\right. \\ \left.f_Q
   \left(\frac{2b}{r-b}+
   \frac{(2r-b)(b'r-b)}{(r-b)^2}\right)  
   +f\frac{r^3}{r-b}\right]\\
   =\frac{1}{2r^3}\left[2rf_{QQ}Q'b+f_Q
   \left(2b+\frac{(2r-b)(b'r-b)}{r-b} \right )
   +fr^3  \right ],
\end{multline}
where $f_Q=d\alpha e^{\alpha Q}$ and $f_{QQ}=
d\alpha^2e^{\alpha Q}$.  Observe that near the 
throat, where $b(r_0)=r_0$, the fractional part 
of the second term on the right-hand side 
dominates, making the other terms negligible.  
As a result, 
\begin{equation}
   8\pi\rho(r)\approx\frac{1}{2r^3}
   (d\alpha e^{\alpha Q})
   \frac{(2r-b)(b'r-b)}{r-b}.      
\end{equation}
Solving for $b'(r)$, we get
\begin{equation}
   b'(r)=\frac{1}{\alpha}(8\pi)\rho(r)
   \frac{2r^2}{de^{\alpha Q}}
   \frac{r-b}{2r-b}+\frac{b}{r}
\end{equation}
and
\begin{equation}\label{E:general}
   b(r)=\frac{1}{\alpha}\int^r_{r_0}
   (8\pi)\rho(r')\frac{2(r')^2}{de^{\alpha Q}}
   \frac{r'-b}{2r'-b}dr'+\int^r_{r_0}
   \frac{b(r')}{r'}dr'+r_0.
\end{equation}

We saw earlier that the energy density $\rho(r)$ 
can be quite small, based on our discussion of 
noncommutative-geometry and Casimir wormholes.
To see the significance of using a small positive 
$\alpha$, let us return to the linear form 
$f(Q)=\alpha Q+\beta$.  It is shown in Ref. 
\cite{pK23} that for the noncommutative-geometry 
case, Eq. (\ref{E:rho}), the linear form 
$f(Q)=\alpha Q+\beta$, with $\beta=0$, yields

\begin{equation}\label{E:shapefunction}
   b(r)=\frac{1}{\alpha}\frac{m\sqrt{\gamma}}{\pi^2}
   \left[\frac{\text{tan}^{-1}\frac{r}{\sqrt{\gamma}} }{2\sqrt{\gamma}}
   - \frac{r}{2(r^2+\gamma)}-                 
   \frac{\text{tan}^{-1}\frac{r_0}{\sqrt{\gamma}} }{2\sqrt{\gamma}}
   +\frac{r_0}{2(r_0^2+\gamma)}
    \right] +r_0. 
\end{equation}
Thanks to the free parameter $\alpha$ from $f(Q)$ 
gravity, the mass of the wormhole, $m(r)=
\int^r_{r_0}\rho(r')4\pi (r')^2\,dr'
=\frac{1}{2}b(r)$ from Eq. (\ref{E:E1}) can now 
be macroscopic.  The point is that this conclusion 
is also valid for the general case, Eq. 
(\ref{E:general}), due to the assumption that 
$\alpha$ is sufficiently small.  So by invoking
$f(Q)$ modified gravity, we have shown that low 
energy-density wormholes can be macroscopic.

\section{Summary}

This paper discusses viable models for macroscopic 
wormholes characterized by a low energy density.  
Such wormholes may not have a sufficiently large 
mass to exist on a macroscopic scale.  However, 
Morris-Thorne wormholes are likely to be compact 
stellar objects, akin to neutron stars, and would 
normally be quite massive.  By invoking $f(Q)$ 
modified gravity, it is shown that the resulting 
extra degrees of freedom enable us to overcome 
these obstacles, thereby allowing certain 
wormholes to be sufficiently massive despite 
the low energy densities.  Particular attention 
is paid to two important special cases, wormholes 
based on a noncommutative-geometry background and 
wormholes whose energy violation is due to the 
Casimir effect.

\end{document}